\documentclass[twocolumn,showpacs,preprintnumbers,nofootinbib,amsmath,amssymb,floatfix,aps]{revtex4-1}
\usepackage{graphicx}
\usepackage{isotope}

\newcommand{\ve}[1]{\ensuremath{\mathbf{#1}}}
\newcommand{\n}[1]{\ensuremath{|\mathbf{#1}|}}
\newcommand{\Ep}{\ensuremath{E_N}}

\newcommand{\sNu}{\ensuremath{\sigma_\textrm{tot}(\nu)}}
\newcommand{\sAnu}{\ensuremath{\sigma_\textrm{tot}(\bar\nu)}}

\graphicspath{{plot/}}

\begin{document}

\title{Neutron knockout in neutral-current neutrino-oxygen interactions}
\author{Artur M. Ankowski}
\altaffiliation{Present address: Department of Physics, Okayama University, Okayama 700-8530, Japan}
\email{Artur.Ankowski@roma1.infn.it}
\affiliation{INFN and Department of Physics,``Sapienza'' Universit\`a di Roma, I-00185 Roma, Italy}
\author{Omar Benhar}
\email{Omar.Benhar@roma1.infn.it}
\affiliation{INFN and Department of Physics,``Sapienza'' Universit\`a di Roma, I-00185 Roma, Italy}

\date{\today}%

\begin{abstract}
The ongoing and future searches for diffuse supernova neutrinos and sterile neutrinos carried out with large water-Cherenkov detectors require a precise determination of the backgrounds, especially those involving $\gamma$ rays. Of great importance, in this context, is the process of neutron knockout through neutral-current scattering of atmospheric neutrinos on oxygen. Nuclear reinteractions of the produced neutron may in fact lead to the production of $\gamma$ rays of energies high enough to mimic the processes of interest. In this article, we focus on the kinematical range suitable for simulations of atmospheric-neutrino interactions and provide the neutron-knockout cross sections computed using the formalism based on the realistic
nuclear spectral function. The role of the strange-quark contribution to the neutral-current axial form factor is also analyzed. Based on the available experimental information, we give an estimate of the associated uncertainty.

\end{abstract}

%


\maketitle

The detection of the antineutrino burst from the 1987A core-collapse supernova in the Large Magellanic Cloud by three independent experiments~\cite{ref:Kamiokande_SN1987A,ref:IMB_SN1987A,ref:Baksan_SN1987A} marked the dawn of a~new era in observational astronomy. That measurement, totaling 24 events, was feasible owing to not-too far distance from the collapsing star and to the extreme nature of supernova explosions. While the gravitational energy released in the act of collapse is $\sim$200--300 times higher than that produced by the Sun over its entire lifetime, $\sim$99\% of it is radiated over a~time scale of a~few tens of seconds in the form of an immense flux of low-energy neutrinos~\cite{ref:Burrows90}.

Supernovae have long been recognized as unique laboratories to study a number of fundamental physics issues~\cite{ref:Raffelt}. The scarcity of the available data seems, however, to be an insuperable problem, because the frequency of the core-collapse events within our Galaxy is estimated to be $1.9\pm1.1$ per century~\cite{ref:SN_rate}.

On the other hand, during the lifetime of the Milky Way, those phenomena have occurred approximately 100 million times, and the Universe witnesses them at the rate of approximately one per second~\cite{ref:Burrows_nature}. All the past core-collapse supernovae have contributed to a diffuse supernova-neutrino (DSN) flux, which is expected to be tiny but continuous in time. Its detection may provide a great deal of information, complementary to that from the future neutrino bursts.

Although its detailed features show some model depen{\-}dence, it is rather well established that the predicted DSN spectrum has a~peak at the value of 4--7 MeV with an exponential drop at higher energy~\cite{ref:Lunardini}. In the low-energy region, $E_\nu\lesssim8$--12 MeV, the DSN signal is not accessible owing to an overwhelming flux of reactor $\bar\nu_e$. On the other hand, in the high-energy region, $E_\nu\gtrsim30$--40 MeV, it is covered by the $\nu$ and $\bar\nu$ flux of atmospheric origin~\cite{ref:Lunardini,ref:Beacom_PRL,ref:Beacom_AnnuRev}. Moreover, at $E_\nu\lesssim16$ (19) MeV, the solar neutrino
flux from the $\isotope[8]{B}$ (hep) chain dominates over the DSN flux. Therefore, in the search for the DSN signal, the energy window $19\lesssim E_\nu\lesssim30$ MeV plays a~pivotal role, and studies aimed at extending this range toward lower values are of paramount importance.

In this context, water-Cherenkov detectors are of special significance. The recent result of the Super-Kamiokande (SK) Collaboration~\cite{ref:Super-K_new}, performed for $E_\nu>17.3$ MeV, has reached the sensitivity comparable to state-of-the-art theoretical predictions.

While neutrinos and antineutrinos of all flavors are present in the DSN flux, the dominant reaction process at this kinematics is $\bar\nu_e$-induced inverse $\beta$ decay of
free protons in the water molecule~\cite{ref:Fogli},
\begin{equation}
\label{eq:DSN_signal}
\bar\nu_e+p\rightarrow e^++n,
\end{equation}
with the rate of a~few events per year in the 22.5 ton fiducial volume of the SK detector~\cite{ref:Super-K_new}.

When this figure is compared to $\sim$2 cosmic-ray muons penetrating the detector every second and $\sim$25 solar-neutrino and atmospheric-(anti)neutrino events identified every day~\cite{ref:Super-K_new}, it clearly appears that very good understanding of the backgrounds is a~prerequisite for the measurement of the DSN signal and for lowering the minimal neutrino energy accessible in the data analysis.

At the current stage, neutrons cannot be detected in SK, and the observation of process~\eqref{eq:DSN_signal} relies on the observation of the positron. Because Cherenkov
detectors do not distinguish $e^+$'s from $e^-$'s nor, at the energy of interest, from $\gamma$ rays, their sources give rise to the backgrounds. The most important of them is the process of oxygen spallation induced by an interaction with cosmic-ray muons, which currently determines the low-energy threshold of the analysis~\cite{ref:Super-K_new}.

The gadolinium doping program at SK~\cite{ref:Super-K_Ga}, based on the idea originally proposed by the authors of Ref.~\cite{ref:Beacom_PRL} to reduce the atmospheric $\nu_e$ and $\nu_\mu$ backgrounds, is
specifically designed to overcome the problem of the spallation.
The DSN signal~\eqref{eq:DSN_signal} will be identified as a prompt positron detection in coincidence with the delayed 8 MeV $\gamma$-ray cascade produced by the neutron capture on the gadolinium nucleus. As a result, the low-energy threshold of the analysis will be dramatically lowered~\cite{ref:Super-K_new}, down to the region dominated by
reactor
$\bar\nu_e$'s. However, the identification of the signal will require that background events involving neutrons produced through mechanisms other than
reaction~\eqref{eq:DSN_signal} and leading to $\gamma$-ray emission are under control at the quantitative level.

The GEANT-based simulations~\cite{ref:GEANT,ref:Ueno} performed for neutrons of energy from 4 MeV to 1 GeV show that their propagation in water can yield cascades of $\gamma$ rays, the spectrum of which is similar to that expected for the DSN signal~\cite{ref:Super-K_new}. It is important to note that neutrons at such kinematics are known to be knocked out from oxygen nuclei by atmospheric neutrinos, of energy extending from a~few MeV to arbitrary high values~\cite{ref:Barr,ref:Battistoni,ref:Honda}.

In this article, we discuss neutron production in the aftermath of neutral-current (NC) interactions of both neutrinos and antineutrinos with oxygen. Covering a broad kinematical range, we provide the corresponding cross sections, obtained within the impulse approximation (IA) approach. These results are of immediate relevance to the ongoing DSN program carried out by the SK Collaboration~\cite{ref:Super-K_new}, as well as to the search for sterile neutrinos being conducted in the T2K experiment~\cite{ref:Ueno}.

The basic assumption underlying the IA scheme is that nuclear interactions can be seen as the incoherent sum of interactions between the beam particles and individual nucleons. The IA-based framework has proven successful in extensive analyses of the large set of data collected by electron-scattering $(e,e^\prime p)$ experiments, in which the knocked-out proton is detected in coincidence with the outgoing electron~\cite{ref:Boffi_PhysRep,ref:Boffi,ref:Omar_RMP}.

We confine our considerations to elastic scattering on quasifree nucleons bound in a~nucleus, customarily referred to as quasielastic (QE) scattering.

Note that the final states involving a~knocked-out nucleon and no pions may also result from the reaction mechanisms analyzed in Refs.~\cite{ref:Serot,ref:Alvares-Ruso} for the carbon target, such as production of the $\Delta$ resonance subsequently decaying into a nucleon and a $\gamma$ ray. Nevertheless, because the associated cross sections are shown to be lower by 2 orders of magnitude, those reactions have not been taken into account here.

The process of NC QE interaction is sensitive to the strange content of the nucleon. The available experimental evidence~\cite{ref:G0,ref:HAPPEX} suggests that the strange contributions to the vector form factors are small and may be neglected in the context of this study.

This is, however, not the case for the strange axial form factor $F_A^s$. In the NC QE axial form factors of proton and neutron, defined as
\begin{equation}\label{eq:FA_NC}
\mathcal{F}_A^p=\frac{1}{2}\big(F_A+F_A^s\big),\qquad \mathcal{F}_A^n=-\frac{1}{2}\big(F_A-F_A^s\big),
\end{equation}
respectively~\cite{ref:Alberico}, the dominant term is the the charged-current (CC) QE axial form factor $F_A$, while $F_A^s$ introduces the opposite-sign corrections. From this fact it clearly follows that the strange axial form factor plays an important role in our considerations, since it drives the asymmetry between the proton and neutron NC QE cross sections.

It is customary to apply the dipole parametrization of $F_A^s$, by analogy to $F_A$, and to use as a~cutoff parameter the same value of the axial mass $M_A$,
\begin{equation}
F_A^s=\frac{g_A^s}{(1+Q^2/M_{A}^2)^2},\qquad F_A=\frac{g_A}{(1+Q^2/M_A^2)^2}.
\end{equation}
This choice is supported by the result from the Brook{\-}haven National Laboratory Experiment 734 (BNL E734), which analyzed proton knockout by (anti)neutrino NC QE interaction using a carbon-dominated target~\cite{ref:BNL_E734_NC}. From the shape analysis of the obtained $Q^2$ event distributions, the BNL E734 Collaboration has found that $\mathcal{F}_A^p=\frac12F_A(1+\eta)$ with $\eta=0.12\pm0.07$, which translates into the strange axial coupling $g_A^s=-0.15\pm0.09$. Rather consistent values of $g_A^s$ have been obtained in the subsequent reanalyses~\cite{ref:Garvey,ref:BNL_E734_reanalysis,ref:Pate_PRL,ref:Pate_PRC} of the BNL E734 data.

The axial coupling constant $g_A$ can be precisely extracted from neutron beta-decay measurements. In numerical calculations, we apply the state-of-the-art value $g_A=-1.2701$~\cite{ref:PDG2012}.

Both $g_A$ and $g^s_A$ are strictly related to the spin structure of nucleon. In the naive quark model, those quantities have the~simple interpretation~\cite{ref:Alberico,ref:Leader,ref:Bass,ref:Aidala}
\begin{equation}
g_A=\Delta d-\Delta u,\qquad g^s_A=\Delta s,
\end{equation}
where $\Delta q$ is the amount of proton spin carried by quarks and antiquarks of flavor $q$ ($q=u$, $d$, $s$).

The information on the spin composition of nucleon is accessible in polarized deep-inelastic-scattering (DIS) experiments.
From the inclusive data collected using a  muon beam and \isotope[6][3]{Li}D target, the COMPASS Collaboration~\cite{ref:COMPASS_inclusiveDIS} has recently extracted the value
\begin{equation}\label{eq:Delta_s_COMPASS}
\Delta s=-0.08\pm0.01\textrm{(stat)}\pm0.02\textrm{(syst)},
\end{equation}
in excellent agreement with the HERMES result from the positron scattering off the deuteron~\cite{ref:HERMES_inclusiveDIS}.
Note that the method of extraction of $\Delta s$ assumes validity of SU$(3)_f$ symmetry in hyperon beta decays. Should SU$(3)_f$ be broken by 20\%, the maximal amount not excluded by the constraints from the KTeV experiment~\cite{ref:KTeV,ref:SU3breaking}, the value of $\Delta s$ would shift by $\pm0.04$~\cite{ref:COMPASS_inclusiveDIS}.

The strange quark contribution to the proton spin can, in principle, be determined more directly from semi-inclusive polarized DIS measurements, in which in addition to the scattered lepton, produced charged pions or kaons are detected~\cite{ref:Aidala}. However, the analysis of semi-inclusive data performed by the COMPASS Collaboration~\cite{ref:COMPASS_SIDIS} leads to the conclusion that $\Delta s$ may be dominated by contributions coming from the kinematic region corresponding to Bjorken $x<0.004$, not covered by current experiments. This issue is the main source of uncertainty associated with semi-inclusive $\Delta s$ measurements.

When all uncertainties are taken into account, the values of $\Delta s$ obtained from inclusive and semi-inclusive polarized DIS experiments appear to be compatible~\cite{ref:DeFlorian}.

We acknowledge that the value $g^s_A$ has been also reported from the MiniBooNE experiment~\cite{ref:MiniB_NC}, which used the Che{\-}ren{\-}kov detector filled with mineral oil, CH$_2$. From the amount of single-proton events in the total NC QE event sample of high kinetic energy, $g_A^s=+0.08\pm0.26$ has been obtained. Note that while being positive, the MiniBooNE-determined value remains in agreement with other discussed measurements within its uncertainty.

In the IA regime, the nuclear cross sections are factorized. For neutrino NC QE scattering off a nucleus, one finds
\begin{equation}\label{eq:xsec}
\sigma_{\nu A}= \sum_{N=p,\,n}\int d^3p\,dEP_N(\ve p,E) \frac{M}{\Ep}\sigma_{\nu N},
\end{equation}
where $\sigma_{\nu N}$ is the elementary cross section and $\Ep=\sqrt{M^2+\ve p^2}$ is the struck nucleon's energy.
In the above equation, the information
on nuclear structure is contained in the nuclear spectral function, $P(\ve p,E)$, yielding the probability distribution of removing  a nucleon with
momentum $\ve p$ from the nuclear ground state, leaving the residual $(A-1)$-particle system with excitation energy $E$.  Accurate estimates of the
nuclear cross sections require calculations of the spectral functions based on realistic dynamical models.

Many important features of nuclear structure can be explained within the shell model, based on the assumption that nucleons behave as independent particles subject to a mean field. Within this approach, in the nuclear ground state, the nucleons occupy the $A$ lowest-energy eigenstates of the Hamiltonian, and the spectral function can be conveniently expressed as
\begin{equation}
P_N(\ve p,E) = \sum_{\alpha} |\phi_\alpha|^2 \delta(E-E^N_\alpha),
\label{eq:P_SM}
\end{equation}
where the sum runs over all occupied states and $\phi_{\alpha}=\phi_{\alpha}(\ve p)$ is the momentum-space wave function associated with
the $\alpha$th eigenstate, belonging to the eigenvalue $E^N_\alpha$.

Nucleon knockout experiments, while confirming the validity of the shell model, have unambiguously shown its limitations.
Owing to strong correlations, nucleons are excited to states of higher energy and the occupation of the shell-model states is reduced to values sizably
less than 1~\cite{ref:Saclay,ref:Nikhef,ref:JLab}. Moreover, their energy distribution acquires finite widths, reflecting the finite lifetime of
single-particle states \cite{ref:Fantoni,ref:Brown&Rho}.

Correlation effects can be accounted for rewriting the spectral function in the form
\begin{equation}
P_N(\ve p,E) = \sum_{\alpha} n_{\alpha} |\phi_\alpha|^2 f_\alpha(E-E^N_\alpha) + P^N_{\rm corr}(\ve p,E),
\label{eq:P_full}
\end{equation}
where $n_\alpha < 1$, and the function $f_\alpha(E-E^N_\alpha)$ has a width which increases with increasing $E^N_\alpha$. The correlation component $P^N_{\rm corr}(\ve p,E)$
is characterized by a distinctive energy dependence. It provides a smooth background, extending to large values of $\n p$ and $E$, and does not exhibit the poles associated
with the complex energies of single-particle states.

The results of theoretical calculations carried out within highly realistic {\em ab initio} approaches clearly indicate that the
momentum distribution,
\begin{equation}
n_N(\ve p) =\int dE P_N(\ve p,E),
\end{equation}
becomes independent of $A$ at $\n p\gtrsim300$~MeV~\cite{ref:Omar_RMP} and therefore that the correlation component of the spectral function is largely unaffected by surface and shell effects. As a~consequence, $P^N_{\rm corr}(\ve p,E)$ of a~nucleus can be calculated  in the local-density approximation (LDA) from the results for uniform nuclear matter~\cite{ref:Omar_NM} of constant density $\rho$,
\begin{equation}\label{eq:P_corr}
P^N_{\rm corr}(\ve p,E) =\int dR\:\rho(R)P^{NM,\,N}_{\rm corr}(\rho,\ve p,E).
\end{equation}

In this article, we use the realistic spectral function of oxygen obtained by the authors of Refs.~\cite{ref:Omar_LDA,ref:Omar_oxygen} in the LDA scheme. It consistently combines the shell structure deduced from experimental $(e,e'p)$ data~\cite{ref:Saclay} with the correlation component determined as in Eq.~\eqref{eq:P_corr}.

The numerical results discussed in this work can be readily reproduced, employing the expression of the elementary NC QE cross section $\sigma_{\nu N}$ for scattering
on an off-shell nucleon explicitly given in our previous article~\cite{ref:gamma}. We apply the state-of-the-art parametrization of the electromagnetic form factors of Refs.~\cite{ref:Kelly,ref:Riodan} and set the axial mass to the value $M_A=1.2$~GeV, determined by the K2K Collaboration~\cite{ref:Gran} using oxygen as a target.

Note that setting $M_A$= 1.03~GeV~\cite{ref:Meissner} would lead to a sizable decrease of our total NC QE cross sections.
For example, at probe energy 0.6~GeV, the change would amount to 13.6\% and 16.3\% for $\nu$'s and $\bar\nu$'s, respectively.
This effect would translate into a 12.1\% and 12.6\% change of the neutron-knockout cross sections for neutrinos and
antineutrinos, respectively.

Increasing $M_A$ with respect to the value extracted from deuterium measurements~\cite{ref:Meissner,ref:BBA03} has been interpreted as an effective way of taking into account multinucleon reaction mechanisms~\cite{ref:Nieves_PLB}. Owing to a~lack of experimental data for neutrino QE scattering off oxygen, the accuracy of such a~phenomenological method cannot be directly verified. However, based on the experience gained in electron scattering, nuclear effects in oxygen are expected to be similar to those in carbon, for which a~consistent analysis of NC and CC QE data has been performed~\cite{ref:carbon}. In Ref.~\cite{ref:carbon}, the approach applied in this paper was shown to yield results in good agreement with the total CC QE cross sections extracted by NOMAD~\cite{ref:NOMAD}, providing a fairly
good description of the NC QE differential cross sections $d\sigma/dQ^2$ measured for neutrinos and antineutrinos by BNL E734~\cite{ref:BNL_E734_NC}. Very good agreement has been found with the shape of the neutrino NC QE differential cross section $d\sigma/dQ^2$ obtained from MiniBooNE~\cite{ref:MiniB_NC}. Hence, our approach is expected  to yield results of comparable accuracy for oxygen.

In our calculations, the strange axial coupling constant is
\begin{equation}\label{eq:gAs}
g^s_A =-0.08\pm0.05.
\end{equation}
We have estimated the overall uncertainty of $g^s_A$ adding in quadrature the errors of the COMPASS measurement, see Eq.~\eqref{eq:Delta_s_COMPASS}, and the uncertainty related to possible SU(3)$_f$ breaking effects~\cite{ref:COMPASS_inclusiveDIS}.
Note that within its uncertainty, the value \eqref{eq:gAs} is in agreement with a~broad class of theoretical predictions~\cite{ref:Wakamatsu,ref:CloudyBag} and that next-generation measurements should be able to determine the value of $g^s_A$ with higher precision~\cite{ref:EIC,ref:EIC_whitePaper}.

\begin{figure}
\includegraphics[width=0.80\columnwidth]{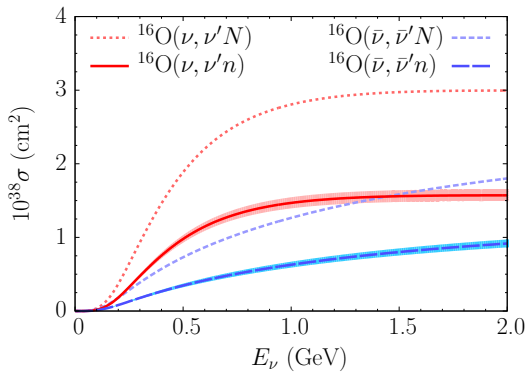}
\caption{\label{fig:cs}(color online) Cross sections for neutron knockout from the oxygen nucleus by NC QE interaction of neutrino (solid line) and antineutrino (long-dashed line). For comparison, the total NC QE cross sections for $\nu$ (dotted line) and $\bar\nu$ (short-dashed line) scattering are also shown.}
\end{figure}

\begin{figure}
\includegraphics[width=0.80\columnwidth]{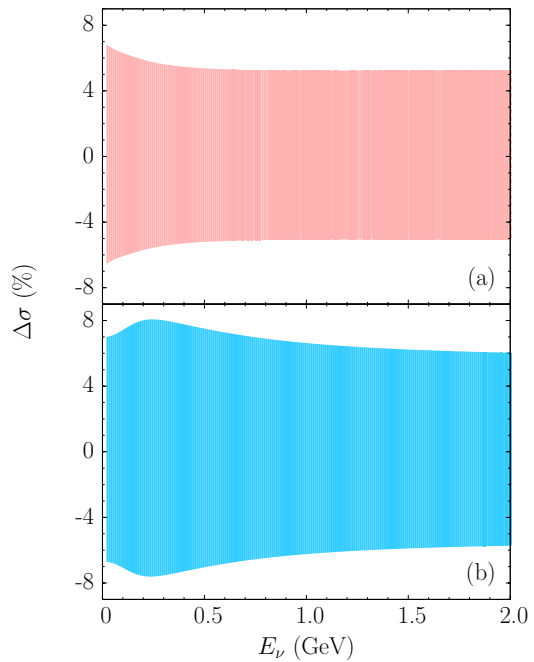}
\caption{\label{fig:uncertainty}(color online) Strangeness-related uncertainty of the cross section for neutron knockout by (a) neutrino and (b) antineutrino NC QE scattering off oxygen.}
\end{figure}

Figure~\ref{fig:cs} shows the calculated cross sections for neutron knockout induced by neutrino and antineutrino NC QE interaction with the oxygen nucleus. The obtained results exhibit similar energy dependence to that of the total NC QE cross sections,  $\sNu$ and $\sAnu$. The reason of such behavior is twofold. First, in the IA, underlying our approach, the gross features of the nuclear cross sections follow from the properties of the elementary cross sections. Second, the elementary NC QE cross sections for scattering off the proton and neutron are largely similar.

The uncertainty of the strange axial coupling constant~\eqref{eq:gAs} introduces to our predictions uncertainties represented by the bands in Fig.~\ref{fig:cs}. To provide insight
into their energy dependence, in Fig.~\ref{fig:uncertainty}, we also show the relative uncertainties, defined as
\[\Delta \sigma=\frac{\sigma(g^s_A)}{\sigma(-0.08)}-1.\]

In the interval $0.2\leq E_\nu\leq 10$ GeV, the cross section for neutron knockout by neutrino (antineutrino) NC QE scattering decreases by less than 5.71 (7.62)\% when the value of $g_A^s$ is fixed to $-0.13$ instead of $-0.08$ and increases by less than 5.90 (8.08)\% when $g_A^s=-0.03$.

Unlike the cross sections for neutron knockout, $\sNu$ and $\sAnu$ exhibit weak dependence on the strange axial form factor. This behavior is easy to understand. In the idealized
case of a~completely isoscalar target, $P_n(\ve p,E)=P_p(\ve p,E)$, the nonvanishing combinations of the form factors are isoscalar-isoscalar, such as $(F_A^s)^2$ and $F_A^s(G^p_M+G^n_M)$, and isovector-isovector, e.g., $(F_A)^2$ and $F_A(G^p_M-G^n_M)$, with $G^N_M$ being the nucleon magnetic form factor. However, in view of the relations $g_A^s\ll g_A$ and $(G^p_M+G^n_M)<(G^p_M-G^n_M)$, it clearly appears that the isoscalar-isoscalar contributions are much smaller than the isovector-isovector ones. To a~good approximation, that picture applies also to the oxygen nucleus, because the difference between its neutron and proton spectral functions is not large~\cite{ref:gamma}, playing an important role only at low energy. For example, in the range $0.2\leq E_\nu\leq 10$ GeV, the total cross section for neutrino (antineutrino) NC QE scattering increases by less than 0.46 (1.57)\% when $g_A^s=-0.13$ is applied instead of the value $-0.08$. The somewhat larger dependence of $\sAnu$ on the strange axial coupling is a~consequence of its large sensitivity to the size of the axial-magnetic term, resulting from the destructive interference between the response functions in the antineutrino cross section~\cite{ref:Amaro_interference}.

\begin{figure}
\includegraphics[width=0.80\columnwidth]{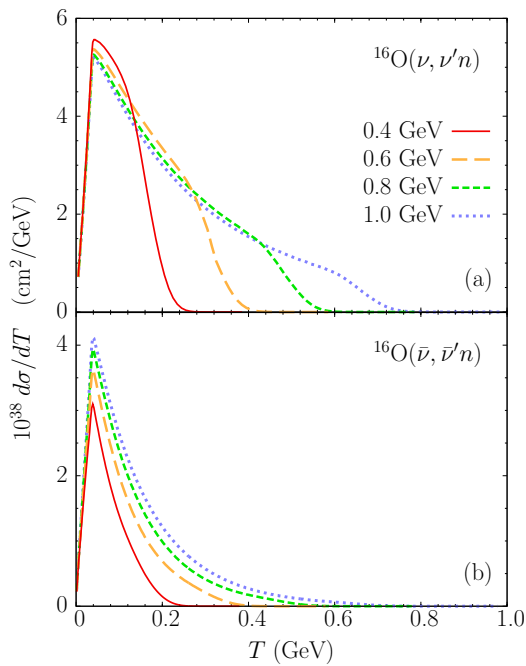}
\caption{\label{fig:diffCSs}(color online) Total kinetic energy distribution of the nucleons knocked out from oxygen in (a) neutrino and (b) antineutrino NC QE interactions with a neutron.}
\end{figure}

An important consequence of that interference is the strong quenching of high energy transfers in antineutrino QE scattering. This feature is illustrated in Fig.~\ref{fig:diffCSs}, showing a comparison of the differential cross sections $d\sigma/dT$, where $T$ is the total kinetic energy of the knocked-out nucleons,  calculated for neutrino and antineutrino NC QE interactions with neutrons bound in oxygen, at fixed energies of 0.4, 0.6, 0.8, and 1.0 GeV.
Note that $d\sigma/dT$ is not expected to be strongly affected by final-state interactions~\cite{ref:MiniB_NC}.

To facilitate use of the total cross sections discussed in this article, in Supplemental Material~\cite{ref:tables}, we provide them in a~tabulated form for energies up to 10 GeV.

Finally, we remark that our approach readily applies to the carbon nucleus~\cite{ref:Omar_NC,ref:carbon}, the significance of which follows from an extensive use of hydrocarbons in neutrino detectors performing also astrophysical searches~\cite{ref:MiniB_SN,ref:NOvA,ref:KamLAND}. In particular, liquid scintillators, while being materials rich in free protons, allow identification of the reaction~\eqref{eq:DSN_signal} by the correlated signal from the prompt positron and the delayed 2.2 MeV $\gamma$ ray originating from the neutron capture on the proton. The lack of reliable calculations of the NC cross sections is in fact one of the main sources of the background uncertainty, which is regarded as the most important limitation
of experimental searches~\cite{ref:KamLAND}.

In addition to the mechanisms taken into account in our work, a~fully comprehensive analysis should also include an estimate of the backgrounds arising from processes in which pions from deexcitation of nucleon resonances are absorbed. A discussion of
the impact of these processes on NC neutrino-nucleus cross sections at intermediate energies can be found, e.g., in Ref.~\cite{ref:Mosel}.

In this article, we have argued that the formalism based on realistic spectral functions is suitable to carry out accurate calculations of the neutron-knockout
cross sections in the kinematical regime of atmospheric-neutrino interactions.
The numerical results of our work cover a broad energy range, extending to 10 GeV, relevant to Monte Carlo simulations.
Using the available experimental information, we have estimated the uncertainty arising from the strange-quark contribution to
the NC QE axial form factor
 and found it to be smaller than reported elsewhere~\cite{ref:Omar_NC, ref:Meucci_NC, ref:Barbaro_NC, ref:Sobczyk_NC}.
As a final remark, we want to emphasize that the results of our work are of immediate use in the ongoing searches for diffuse supernova neutrinos and for sterile neutrinos.

\begin{acknowledgments}
We would like to thank Makoto Sakuda for drawing our attention to the importance of the subject covered in this article, as well as for a critical reading of the manuscript. A most informative conversation with Paolo Lipari on atmospheric neutrinos is also gratefully acknowledged. This work was supported by the INFN under Grant No. MB31.
\end{acknowledgments}

\end{document}